# Privacy protection and service evaluation methods for location-based services in edge computing environments

Shuang Liu

College of Information Science and Engineering, Hunan University,410082,ChangSha Chnia

lshuang@hnu.edu.cn

**Abstract**.
This paper proposes a privacy protection and evaluation method for location services based on edge computing environment. By constructing the site service data protection and system evaluation system in the edge computing environment, based on the existing user privacy protection work, the data processing module and service evaluation module are constructed, and the evaluation algorithm is designed. NPE evaluation model and POE evaluation model are designed according to relevant research recommended by IPE. Specifically, in the NPE evaluation model, we regard each user's decision as a group of factors, and propose a method to integrate learning factors. In the poe evaluation model, users' hidden intentions for the next action are understood by unifying metadata information, two time contexts and other different factors. The experiment verifies the effectiveness and feasibility of this method.

## 1 Introduction

With the advent of the Internet of Things and the proliferation of mobile devices, location-based social networking (LBSN) has penetrated people's lives. People can access the Internet through mobile devices such as cell phones, smartwatches, and smart cars, and use geographic information and social attributes in LBSN to determine users' geographic preferences when visiting points of interest (POIs). However, the current explosion of Internet applications and users has led to significant exploitation of the potential of mobile edge computing. LBSN-based POI recommender systems must first gather information about users and then provide them with content that may be of interest to them [1]. However, providing recommendations inevitably involves data about users, and the paths contained in user data may reveal sensitive information about users. Potential attackers can effectively use users' private information obtained from recommendation results. On the one hand, mobile edge computing infrastructure is often deployed at the edge of the network, e.g., in wireless base stations, which makes it more vulnerable to insecure environments [1]. On the other hand, mobile edge computing will use open application programming interfaces, which makes edge servers easily vulnerable to external attackers. Due to the limited resources and capabilities of edge devices, mobile edge computing services are exposed to certain security risks.

Therefore, user privacy needs to be taken into account while improving the accuracy of the POI recommender system. If POI recommender systems are equipped with privacy mechanisms to protect sensitive user data, this may affect the accuracy of recommendations and thus reduce user comfort. Currently, there is no uniform evaluation standard for POI [2] recommendations in edge environments, and the quality of POI recommendations varies.

To improve the quality of location-based POI recommendations while protecting users' privacy, this chapter proposes a novel approach for privacy protection [3] and service evaluation based on edge computing environments. On the one hand, users' sensitive information is managed by using RSA encryption algorithm and processing pseudonyms for users, and on the other hand, k-anonymity is used to prevent users from being distinguished. On the other hand, the recommendation results of POIs returned by the server are evaluated by developing the NPE and POE evaluation models under the protection of global privacy mechanisms. The NPE evaluation model analyzes users' decisions based on their choices and can identify the key factors of their POI selection decisions by considering each user's decision as a set of factors and providing a method for integrating learning factors. An innovation of the model is that it can identify the key factors while preserving the user's decision structure by maximizing a new scalar projection as an objective. The POE evaluation model takes into account multiple factors related to the POI and the user, and combines different factors in a unified way to learn the hidden intention of the user's next action [3]. Using a linear equalization function, temporal context, sequential relationships, geographic factors and additional metadata information are integrated into an integrated architecture. To further exploit the influence of sequential relationships and geographic location in the context of POI recommendations [4]without feature engineering, we have also developed a method to pre-train POI embeddings in POE.

## 2 Architecture of models for privacy protection and evaluation of location-based services

The computing resources and limited transmission power of mobile devices in a simple edge computing environment are not sufficient to recommend the best points of interest to users[4], and the proximity of edge nodes to users' terminals and the possibility of disconnected connections between edge nodes compared to the cloud could lead to attacks on users' personal data. Processing and computation by cloud servers is also not sufficient to protect users' privacy. A single cloud server may be the only target for an attacker, and sending to that cloud server may lead to increased latency and degrade the quality of service, especially if large amounts of data must be transmitted over an already congested backhaul connection.

This section presents the system model used for the location-based approach for privacy protection and service evaluation in an edge computing environment, describing the data processing module and the result evaluation module accordingly.

The system model for location service privacy in edge computing environments consists of four main components: the edge server, the cloud server, the user, and the trusted LBSP server (client). When a user makes a location service request [5], which contains personal and demand information, such as the address of the service request and the type of service requested; the information is received by the edge server and the cloud server system and evaluated by the LBSP server system; the location service provider (LBSP) is used to receive the location service request sent by the user, distribute it and send it to the edge server and cloud server after processing the data The LBSP servers are used to receive the location service requests from users, process the data, distribute it to the edge and cloud servers, and evaluate the results returned by the edge and cloud servers.

Considering the popularity of edge devices and their increasing computing power, and in order to increase the flexibility and autonomy of the solution, we assume that most of the current services can be quickly queried from the edge server to get the results and save a lot of communication time cost, which enables faster querying under the assumption that the needs of location-based services are basically covered [5]. Of course, not all query results coming from edge servers are reliable and trustworthy, so the client system of this solution evaluates the query results coming from edge servers as soon as it receives them and uses the result evaluation module to determine whether the basic thresholds set by the evaluation model are met. Thus, if the results returned by the edge server do not meet the threshold set by the result evaluation module, the client system forwards the query originally sent to the edge server to a cloud server with higher processing capacity. After the cloud server returns the result, the result is also evaluated by the result evaluation module, and the result is communicated to the user by comparing the evaluation results of the query results returned by the edge server and the cloud server for the same query in the system evaluation module.

The solution can be divided into four parts: The user sends a request, the system processes the data and sends a query, the edge or cloud server queries and provides the results, and the system evaluates and processes the results.

All users who require location-based services and are registered with the system server on the client side must upload information about the space they are in, the time of service they require, and the type of POI to be queried each time they make a service request via their mobile device. The pseudonymization is performed in a timely manner and the location data is compensated accordingly, while the -anonymization method is used to send the individual queries primarily to edge servers within the coverage area of the system.

The solution can be divided into four parts: The user sends a request, the system processes the data and sends a query, the edge or cloud server queries and provides the results, and the system evaluates and processes the results.

All users who require location-based services and are registered with the system server on the client side must upload information about the space they are in, the time of service they require, and the type of POI to be queried each time they make a service request via their mobile device. Timely pseudonym processing is performed to offset the location data accordingly, and the $k$ -anonymization method is used to send $k$ query requests to the edge servers within the system coverage in priority.

**3 Design of a data processing module**

In a location-based privacy protection system for edge computing environments, each user sends a location-based service request, i.e., a POI query [5], where the user's personal information, including location, previous query information, etc., and the privacy of the user's query information are often targeted by attackers. Queries can also be intercepted or manipulated by the LBSP server when it sends certain queries to the edge or cloud server, resulting in privacy violations and affecting the quality of service provided by the edge and cloud servers. As a result, the edge and cloud servers [6] included in the system are often considered to be untrusted third-party servers.

When a user sends a request to a client, the client often chooses to send the user's request to an edge or cloud server in its vicinity, and the edge and cloud server in its vicinity may directly have private information about the user, such as the user's location information and the request content, which affects the user's privacy. To solve these problems, the client needs to process the user's personal information and the service request [6] made by the user before sending the request to the edge server and the cloud server, removing as much sensitive information about the user as possible by using methods such as pseudonyms and sending $k$ queries to the edge servers and cloud servers using $k$-anonymous methods, so that attackers cannot accurately intercept or tamper with the query requests to protect the user's privacy. This protects the user's privacy and ensures that the feedback from the edge and cloud servers is really based on the queries sent by the client.

The RSA encryption algorithm used in the encryption process of which is

(1) Select two large prime numbers $p$ and $q$ ($p \neq q$).

(2) Calculate $N = p \times q$, $\varphi(N) = (p-1) \times (q-1)$ respectively.

(3) Select an integer $e$, $e$ satisfies $0 < e < \varphi(N)$, and $e$ and $\varphi(N)$ are mutually exclusive, thus obtaining the public key $(e, N)$.

(4) For $e \times d \equiv 1 \mod \varphi(N)$, find $d$ by extending Euclid's algorithm to obtain the private key $(d, N)$.

The specific steps for RSA public key encryption and private key decryption are as follows.

(1) Encryption process: known plaintext $m$, public key $(e, N)$, column $m^e \equiv c \mod N$, from which the ciphertext $c$ is solved.

(2) Decryption process: the ciphertext $c$ is known, the private key $(d, N)$, the column $c^d \equiv m \mod N$, and the plaintext $m$ is solved.

Encryption of the private key and decryption of the public key according to RSA (digital signature) is performed as follows.

(1) Encryption process: known plaintext $m$, private key $(d, N)$, column $m^d \equiv c \mod N$, from which the ciphertext $c$ is solved.

(2) Decryption process: the ciphertext $c$ is known, the public key $(e, N)$, the

column $c^e \equiv m \bmod N$, and thus the plaintext $m$ is solved.

The mobile devices receive real-time information about users [7] in the coverage area who will soon be using the service and create their respective databases; wherein, for any user $U_i$, the information includes at least the respective user identification ID and the exact location information loc, and also needs to include the attributes char of the service to be queried; the pseudonym and the required public key [7], private key and certificate are stored in the user's mobile device, and the user $U_i$'s last information related to the point of interest q, including the time t of the last visit of the user $U_i$, etc. When the user $U_i$ makes a service request, the system randomly picks k-1 user locations from the anonymous range and randomly generates k-1 requests, which include the user's pseudonym fID, location loci, and a random query service attribute $char_i$. That is, user fID set fID = {fID1, fID2, ---, fIDk}, location loc = {loc1, loc2, ---, lock}, and service attribute $char$ = {char1, char2, ---, chark}. where the pseudo-identification QT is sensitive familiar and QT = {fID, loc}.

After each user's location service request, the system assigns a random pseudonym to each user and encrypts the user's information using an RSA encryption algorithm that performs a multiplicative homomorphic encryption.

The homomorphic encryption system is able to perform calculations on the encrypted data without knowing the private key. These calculations lead to a result that is itself encrypted. The result of a calculation with encrypted data is the same as the result of a calculation with the original data.

If a customer wants to perform calculations with its data on a cloud or edge server, it must share the key with the provider to decrypt the data [8]. The cloud or edge server accesses the data through the shared key. Therefore, to solve this problem, homomorphic encryption is used. The client allows the cloud or edge server to compute the data without decrypting it [8]. The result is returned to the client and remains encrypted. Since the client side is the sole owner of the key, the other parties cannot decrypt the data and the results.

**4 Design of a data processing module**

The evaluation module includes the NPE evaluation model and the POE evaluation model. The evaluation model NPE is used by integrating the determinants D={f1, f2,..., fn} of the user's point of interest POI, D includes factors such as identity, category, brand, distance and popularity, all denoted by fi. Each factor is embedded in a vector and the maximum value projected by the scalar is used as a key reference.

The POE evaluation model takes into account the user's access history and the purpose of the visit, nests the hidden purpose [9] of the user's visit using a nonlinear activation function, considers factors related to the time and space in which the user u submits the request, and considers other factors that may influence the decision, and calculates the score of the point of interest $q_l$ recommendation at time $t_i$ through the hidden vector of user u, etc.

## 4.1 Evaluation model NPE

We use factors from all aspects to measure the user's decision. This is to avoid missing possible clues. In practice, however, it is unlikely that all factors are included in the decision process. Typically, each decision consists mainly of a few key factors (e.g. 2-5) that we try to identify, while the other supporting factors have secondary influence.

The NPE evaluation model provides the decision D and selects a small subset of factors that positively influence D, thus determining the influence weights of key factors [9]. For example, the factors f1, f2, f3, f4... for the point of interest identifier, category, brand, and popularity of the recommended outcomes are ranked as f1, f2, f3, f4... and the decision factor D for evaluation = {f1, f2,..., fn}, ranking each factor

Embed the factors f1, f2, f3 --- in the determinants D in vectors $f_1$, $f_2$, $f_3$ --- respectively. Assuming that the number of main determinants is 2, find the scalar projection of the sum of any 2 vectors of all factors to maximize i.e. $\max \hat{f}^T d / |d|$.

Where $d$ is the embedding of the aggregation key and $\hat{f} = \sum_{i=1}^{n} f_i$, |-| is the Euclidean parametrization.

Next, let $F = [f_1, f_2, f_3, ... f_n]^T$ be the embedding matrix of the decision factor D and the n-dimensional vector $a$. Formally, the scalar projection maximization objective of the decision factor D is as follows.

$$\max_{F,a} \frac{(\sum_i (a_i f_i))^T \sum_i f_i}{|\sum_i (a_i f_i)|}, a_i \geq 0, nzr(a) \leq \lambda \quad (1)$$

where $nzr(a)$ is the number of non-zero items in $a$ and $\lambda$ is the sparsity threshold that limits the number of key factors. This means that the vector $a$ represents the contribution of the key factors in the aggregated embedding, i.e. $d = \sum_i (a_i f_i)$.

Sparse constraints make the optimization problem difficult. First, it is easy to verify that the optimization is not convex and that equation (1) is still a linear system with sparse constraints, which is usually NP-hard. To efficiently solve this combinatorial optimization problem, machine learning can be used for optimization. Instead of optimizing directly, we predict the probability of each factor becoming critical as the factors are integrated, and these integrations are updated to maximize the scalar projection.

If their maximum value of the scalar projection reaches a systematic threshold, the evaluation continues, otherwise it is considered a failure.

In the actual application of the algorithm for evaluating the determinants in the NPE model, which is also to be optimized, the evaluation is purely data-driven, directly assessing the probability of each factor being a critical factor [10]. However, whether or not a factor is a critical factor depends largely on the scalar projection of other factors onto it [10]. Intuitively, a key factor should be supported and supported by many other factors that have a high scalar projection of that factor. Therefore, we use self-projected attention to compute a different embedding of each factor that encodes the projection information to learn the likelihood in a next step.

（1）Model sparse release estimation function

Formally, given factors f1, f2, f3, ..., fn in a decision and the corresponding embedding matrix of the factors is $F=[f_1,f_2,f_3,...f_n]^T$. The scalar projection matrix P of pairwise occurrences is first computed, where $P_{ij}=f_i^T f_j/|f_i|$ is the scalar projection of the vector $f_j$ on the vector $f_i$, and then P is normalized using row-wise softmax as follows.

$$\hat{P}_{i:} = \text{soft}\max(P_{i:}), i \in \{i,...,n\} \quad (2)$$

After that, it calculates one intent embedding for each factor fi and $\hat{P}_{i:} = \text{soft}\max(P_{i:}), i \in \{i,...,n\}$ is the weighted sum of all factor embeddings $f_j$ obtained through $\hat{P}_{ij}$. For example, as shown in Figure 1, it can be seen that the dizziness information is well encoded in $\hat{f}_1$ therefore has a higher scalar projection factor, i.e., f1 and f2 contribute more to $\hat{f}_1$, which corresponds to a matrix of the form $\hat{F}=\hat{P}F$.

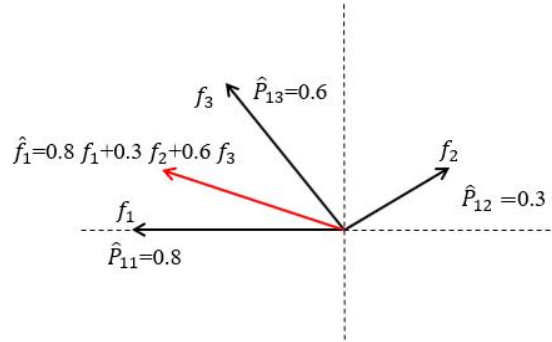

Figure1　Example of embedding vector weighting

Using the output vector and the embedding intention vector, the sparse probability estimator module evaluates the probability that each factor can be used as a key factor. The two factor embedding matrices are first concatenated in $\hat{F} \oplus F$ and fed to a three-layer multilayer perceptron to everywhere n-factor fee-normalized release vectors $l$.

$$L = Dropout(\text{Re}LU((\hat{F} \oplus F)W_1 + b_1)) \quad (3)$$

$$l = Dropout(\text{Re}LU(LW_2 + b_2))W_3 \quad (4)$$

where $M_1 \in R^{d \times d}$, $M_2 \in R^{d \times d}$, $M_3 \in R^{d \times d}$, $b_1 \in R^d$, $b_2 \in R^d$, and $b_3 \in R^d$ are the parameters of the trainable multilayer perceptron. The non-normalized likelihood of factor fi $l_i$ is determined by both $f_i$ and $\hat{f}_i$, That is, the probability of each factor takes into account its own and other factors, focusing on factors with a high projection scale.

Sparse activation is then used to normalize the l. The method considers the use of sparsemax, which is similar to softmax, except that the sparse probabilities of the output:

$$sparse\max(l) = \arg\min_{p \in \triangle^{n-1}} |p - l| \quad (5)$$

where $\triangle^{n-1} = \{p \in R^n | 1^T p = 1, p \geq o\}$ is the n-dimensional monomorphic probability and the sparsity is guaranteed by the Euclidean projection to $\triangle^{n-1}$. In practice, $sparse\max(l)$ can be computed in $O(n \log n)$ time.

Finally $\hat{l} = sparse\max(l)$ is the effective likelihood vector indicating which factors are critical and their influence weights. $\hat{l}$ The number of non-zero entries is not controllable, but it is often necessary to provide flexibility for the number of critical factors, e.g. $\lambda$ in xxx. Therefore, the normalized $l$ before sparsemax is further regulated by adding $|\sum_i f_i|$ to $|\sum_i f_i|$. The higher the weight of , the more key factors are identified by $\hat{l}$. However, $|\sum_i f_i|$ regularized sparse activation is a more relaxed one and does not guarantee that the number of key factors is limited by a certain threshold.

<b>Decision structure learning Given the likelihood vector $\hat{l}$, the aggregated key embedding d of decision D is computed as follows.

$$d = \sum_{i=1}^{n} \hat{l}_i f_i = \hat{l}^T F \quad (6)$$

We then preserve the decision structure by maximizing the sum of the scalar projections of all relevant factors embedded in d:

$$\max_F \hat{f}^T d / |d| \quad (7)$$

In contrast to Eq. (3.1) and $\hat{f}_i = \sum_{i=1}^{n} f_i$, its objective does not require a search vector a.

To learn how to embed factors, both positive and negative decision examples are needed, and while decision D is unambiguous when a user chooses to access a POI in a particular context. Using a representation of empirical access metrics, for each single-valued decision D, we can create multiple negative examples $D^-$ by replacing factors associated with a given POI with factors from other POIs that the user is unsure about visiting. Typical alternative POIs for the negative cases could be those near the visited POIs or in a category related to the visited POIs. Similarly, $V\hat{R}(D^-)=1$.

Prediction of visit rates by scalar projection $V\hat{R}(D)=1$ and $V\hat{R}(D^-)=1$, i.e. $\sigma(\hat{f}^T d/|d|)$, where $\sigma(x)=1/(1+e^{-x})$ is the sigmoid function. $VR(.)$ and $V\hat{R}(.)$ can be viewed as the projected and empirical distribution of decision instance visit rates. The decision structure is then learned by minimizing the distance $dist(VR(.),V\hat{R}(.))$ between the two distributions to retain the decision structure. Replacing dist(-, -) with KL-divergence and ignoring the constants is minimized by

$$O = -\sum_D \log VR(D) - \sum_{D^-} \log(1-VR(D^-))$$

(8)

In other words, the objective sums all decision cases and maximizes $VR(D)$ for positive cases and minimizes $VR(D^-)$ for negative cases. Note that maximizing $VR(D)$ or minimizing $VR(D^-)$ is actually maximizing or minimizing the corresponding scalar projection $\hat{f}^T d/|d|$.

**4.2 Evaluation model POE**

By extracting hidden intentions about the user's next action, which are adaptively modeled with temporal context, user-based and POI-based spatial intentions are extracted separately using nonlinear transformations. Through this non-linear separation and extraction, additional data information such as temporal context is integrated into user-based and POI-based intention learning in an integrated manner. First, the user's next POI recommendation problem is based on a sequence of historical POIs visited by a given user before time $t_{i-1}$ $I_i^u = \{q_{t_1}^u, q_{t_2}^u, ..., q_{t_{i-1}}^u\}$, and the task calculates a score for each POI based on time $t_i$ and $I_i^u$, indicating the probability that the user visited that POI at time $t_i$.

In most location-based information services, in addition to the historical POI access sequence, the user and the POI are associated with ancillary data information. For example, a user may contact other users (e.g., friends, family, classmates, colleagues) with each other to share his activities or opinions, and the POI may also include text descriptions and category tags, with the ancillary data associated with user u represented by $A_u$ and the ancillary data associated with POI q and represented by $A_q$. In building the basic model, the method also introduces an additional neural network layer to model the user's spatial intent and to model the. $u_u \in R^d$ is the embedding of user u, $q_l \in R^d$ is the embedding of the recommended POI candidate, and $q_{t_{i-1}}^u \in R^d$ is the embedding of the POI visited by user u at the time of $t_{i-1}$. The hidden intent of the next visit is modeled by a nonlinear activation function, correcting the linear unit to $\mathrm{Re}LU(x) = \max(x,0)$.

$$d_{t_i}^q = \mathrm{Re}LU(W_1 q_{t_{i-1}}^u + b_1) \qquad (9)$$

$$d^u = \mathrm{Re}LU(W_2 u_u + b_2) \qquad (10)$$

$$c_l = \mathrm{Re}LU(W_3 q_l + b_3) \qquad (11)$$

where is the hidden intent vector is capturing the semantics about the user's next action based on the user's last POI visit at $t_i$, $d^u$ is capturing specific information about the user's preference for a particular space, $c_l$ is the intent of the candidate point of interest. $l$ $M_1 \in R^{d \times d}$ is the transfer matrix from the POI embedding, $M_2 \in R^{d \times d}$ is the transfer matrix from the user embedding. $M_3 \in R^{d \times d}$ is a weight matrix, $b_1 \in R^d$, $b_2 \in R^d$, $b_3 \in R^d$ are the offset vectors.

Then, using the hidden intent vectors $d_{t_i}^q$, $d^u$ and $c_l$, the recommended score for the interest point $q_l$ at time is $t_i$:

$$y_{u,t_i,l} = (d_{t_i}^q + d^u)^T c_l \qquad (12)$$

In summary, the model does not directly use the user and POI embedding vectors, but uses the feed-forward network layer embedding converted into an intent vector, which represents the recommendation score based on the intent vector. Both the transfer matrix and the displacement vector are able to identify the most useful signals in the embedding and can also be extended more easily and intuitively by separating the intent vector from the embedding vector and combining signals from different contexts.

(1) Combining metadata information.

Since the relevant metadata information can provide additional knowledge for the user or the POI, the understanding of the user's motion can be improved by taking $A_u$ and $A_q$ into account. Therefore, the model can be further refined by combining metadata information from different contexts. Therefore, the model can be further refined by including these auxiliary semantics in the intent computation.

By nesting the hidden intent of the user's current visit using a nonlinear activation function, the intent vector $d_{t_i}^q$ attempts to capture the intent with respect to the user's next step by capturing the time of the last visit to the point of interest. the intent vector $d^u$ is to obtain specific information about the user's preference for a particular space, and $c_l$ is the intent of the candidate point of interest $l$. Also considering other relevant determinants to complement the user $u$ and the point of interest $q$, there are cofactors $m_q$ and $m_u$ for:

$$m_q = \frac{1}{|A_q|} \sum_{m \in A_q} m_m \tag{13}$$

$$m_u = \frac{1}{|A_u|} \sum_{m \in A_u} m_m \tag{14}$$

where $m_m$ is the embedding of item m in the metadata $A_u$ and $A_q$, respectively, then:

$$d_{t_i}^q = ReLU(W_1(\alpha q_{t_{i-1}}^u + (1-\alpha)m_{q_{t_{i-1}}^u}) + b_1) \tag{15}$$

$$d^u = ReLU(W_2(\beta u_u + (1-\beta)m_u) + b_2) \tag{16}$$

$$c_l = ReLU(W_3(\alpha q_l + (1-\alpha)m_{q_l}) + b_3) \tag{17}$$

$M_1 \in R^{d \times d}$ is the transfer matrix from POI embedding, $M_2 \in R^{d \times d}$ is the transfer matrix from user embedding. $M_3 \in R^{d \times d}$ is a weight matrix, $b_1 \in R^d$, $b_2 \in R^d$, $b_3 \in R^d$ are offset vectors, and $\alpha$ is an adjustment parameter that controls the importance of metadata information.

It is important to note that the user embedding and the POI embedding need not be in the same hidden space. In this sense, $A_u$ and $A_q$ are of the same type and the model allows flexibility in linking the two sets of metadata information. This is justified because the two sets of metadata can have different meanings. Just as users and POIs can be associated with textual tags, POI tags can be overlaid with associated services, while users indicate their habits and preferred locations through tags. In this case, it makes more sense to use two separate embedding areas.

(2) Contextualization of time

Visits to historical points of interest with different time intervals contain different spatial intentions, so user intent also includes relevant factors such as the time of embedding. and there are two types of available temporal contexts, one is the time interval between two consecutive POI visits ($t_i - t_{i-1}$), and the other is the specific time point of the POI visit ($t_i$). Users may express different spatial intentions at different times of the day. Therefore, a combination of both temporal contexts is required to compute POI intent. The time interval since the user's last visit to a POI is critical in determining the user's next behavior; however, a discrete temporal dimension is not appropriate because it is a continuum. Historical visits to POIs at different time intervals may involve a variety of spatial intentions, and the interaction between intentions and time may also be complex and subtle; therefore, replacing the matrix in equation (9) with a transfer matrix related to time interval t, $W_\pi(t)$

$$\mathbf{W}_\pi(t) = \begin{cases} \dfrac{\pi - t}{\pi} \mathbf{W}_0 + \dfrac{t}{\pi} \mathbf{W}_\pi, & \text{for } t < \pi \\ \mathbf{W}_\pi, & \text{for } t > \pi \end{cases} \quad (18)$$

where $\mathbf{W}_0$ and $\mathbf{W}_\pi$ are the two transfer matrices that are the threshold values for the $\pi$ interval.

For the temporal information accessed, time can be partitioned by associating with each time interval a specific displacement vector $b_t$. The corresponding latent intention vector in equation xxx is expressed as:

$$d_{t_i}^q = \operatorname{Re} LU(W_\pi(t_i - t_{i-1})(\alpha q_{t_{i-1}}^u + (1-\alpha) m_{q_{t_{i-1}}^u}) + b_{t_i}) \quad (19)$$

(3) POI embedding pre-training.

The POE evaluation model can be pre-trained by encoding the sequential relationships between POIs and geographic locations through POI embedding pre-training. Since the objective function is not convex, there is no global optimal solution and the optimization strategy in this case is to find a local optimum. The model uses a neural network learning method, DeepWalk, to learn the embedding of each POI. DeepWalk constructs short sequences of nodes based on random walks in the network structure. The node embeddings are then learned using the SkipGram neurolinguistic model, which uses the probability of maximizing the neighbors of the nodes in the sequence. To preserve both types of information in the potential embedding space, the network structure is constructed so that each point of interest is treated as a separate node in the network. Sequences of random walks over the POIs are created by a mixture of POI movement patterns and geographic factors. The random movements from one POI to another in the network were calculated as follows.

$$p(q_j \mid q_i) = \rho \frac{k(q_i, q_j)}{\sum_k \kappa(q_i, q_k)} (1-\rho) \frac{f_{q_i,q_j}}{\sum_k f_{q_i,q_k}} \qquad (20)$$

$$k(q_i, q_j) = 1/(1 + e^{5\frac{d(q_i,q_j)-d}{\sigma(d)}}) \qquad (21)$$

$\sigma(d)$ are the mean and standard deviation of $d(q_i,q_j)$, respectively, and $f_{q_i,q_j}$ is the frequency of the transition from $q_i$ to $q_j$ in the training data set.

The first term on the right-hand side of equation (20) captures the inherent geographic influence between the points of interest, and the second term captures the transient behavior of the numerous users. a is used to balance the two components. Second, the SkipGram language is used to model over these random wandering sequences. Once the learning of the embedding is completed using SkipGram, pretrained embeddings of interest points are used as initialization for model training.

Alternatively, initial user embeddings can be obtained with pre-trained POI embeddings U. First, the frequency of interest points seen by user u in the training dataset is counted, and then the initial user embeddings are weighted by the normalized frequency of:

$$\mathbf{u}_u = \frac{1}{|\mathbf{L}_u|} \sum_j f_j^u \cdot \mathbf{q}_j \qquad (22)$$

where $|\mathbf{L}_u|$ is the number of times user u visited the POI in the training set, and $f_j^u$ is the frequency of user u visiting the point of interest point $\mathbf{q}_j$

In this section, the algorithmic principle of the evaluation model is described in detail. The results returned by the edge server first enter the NPE evaluation model, which determines the number of determinants based on the POI type, selects the appropriate fi factor, such as the POI identifier, initializes the relevant parameters, such as the determinant D, and sets the threshold Thre1. Before the POE evaluation model begins evaluation, the POE model is initialized to build each POI as a separate node in the network structure. Random walk sequences on POIs are created from a mixture of offset models and geography-based factors. The user embedding $u_u$ is evaluated after using the pre-trained POI embeddings to initialize it according to equation (20).

If the edge server passes the NPE evaluation model, it goes to the next level of the POE evaluation model, and if it passes both evaluation models, it returns the query results directly to the user. If the edge server passes only the NPE evaluation but not the POE, or fails the NPE model evaluation, the system forwards the query to the cloud server.

After receiving the results from the cloud server, the system evaluates them in the same way.

## 5 Experiment

### 5.1 Experimental Setting

The data used in the experiment is Gowalla, which is also one of the most important location-based social networks. The recording stores the login history of all users, including detailed times and location. Gowalla's data includes 1313 users, 2196 POIs and 45410 identifiers. Each user has an average of 34.58 POI points. For Gowalla records, we deleted users with fewer than ten POI records and fewer than ten login users. In the processed Gowalla data, 70% of the samples are used for model preparation training, 20% for validation and 10% for testing.

The experiment also selects two real private car data sets from Shenzhen, China, and extracts one data set, such as verification. The chosen area is (114°040E-114°210E, 22°500N-22°650N), which includes Shenzhen Convention and Exhibition Center, Huaqiang North Electronics City and other hot spots. Data on private cars for two months (1 April 2018–31 May 2018 from 8:00 a.m. to 6:00 p.m. daily) were collected, and some data on parking time and abnormal time and location were omitted.

AUC(Area Under the ROC Curve), recall ratio, F1 (accuracy and recall weighted harmonic mean, accuracy and recall ratio can be considered) and PREC accuracy are used as evaluation indicators in this experiment. Evaluation index chart. In addition, this paper also uses the average precision score index (Mean Average Precision，MAP). At present, the most commonly used indicators to evaluate the quality of detection models are generally used to evaluate the quality of classification. They are usually used as comprehensive evaluation indicators for destination prediction. The higher the value, the better the performance.

We compare our approach with the following reference algorithms that maintain the decision structure:

GE：It is a graph deposition method for recommending sites [11] that use logical return to combine proximity between different devices.

MP2VEC：It's a heterogeneous network embedding method [12]. Taking into account the path, POI-X-user-X-POI..., where X can be any type except user and POI.

### 5.2 Experimental results and analysis

For the ability of the NPE evaluation model to maintain the decision structure, we first assess the overall performance of the method used to distinguish between positive and negative user decisions. The relationship between NPE and decision analysis is that NPE maintains the decision structure by identifying key factors and maximizing scale projection. The effectiveness of the NPE [12] in maintaining the decision-making structure depends directly on the quality of the key factors identified. As examples of positive and negative decisions differ only partially in POI-relevant factors, this task is relatively difficult. Our advantage is to distinguish between key and non-key factors. At ACU, our method also prevails over the possibility of positive cases. In Gowalla, NPE can rank random positive cases higher than random negative cases with a probability of 0.91.

For embedding POI before training, DeepWalk generates POI sequences in the model to code the direct order relationship and geographical impact of POI. The scaling parameter $\rho$ is used to balance the two POI direct geographic influences and the excess behavior component $p(q_j | q_i) = \rho \frac{k(q_i, q_j)}{\sum_k \kappa(q_i, q_k)} (1-\rho) \frac{f_{q_i, q_j}}{\sum_k f_{q_i, q_k}}$.

Table 1 shows the performance of different $\rho$ evaluation models in the dataset. The symbol "–" indicates the PI integration model that failed to initialize the pre training. First, we found that the model with pre formed IPE integration initialization is much better than the model without initialization. By comparing the geographical location and conversion model of the two IPEs, the reliability of IPE pre training integration is verified. Secondly, all parameters of different a (positive) have similar performance. If $\rho$ =0, that is, it is not affected by geography, the performance is better. This indicates that the geographic distance and conversion mode do not contain any additional information, and that users are more likely to access locations near the most recently accessed location when accessing the next location. In this sense, geographic impacts can be encoded with transformation patterns and verified through results. So let's put $\rho$ =0 in the experiment.

Table 1 Performance of the evaluation model at different $\rho$

| $\rho$ | MAP |
| --- | --- |
| - | 0.0711 |
| 0 | 0.1735 |
| 0.3 | 0.1633 |
| 0.5 | 0.1601 |
| 0.7 | 0.1645 |
| 1 | 0.1688 |

As for time context, we first consider the influence of two time contexts. By introducing time intervals, we can obtain more performance gains. In particular, the time interval since the last visit of IPE has played an important role in learning spatial intentions from historical spatial behaviors. It can be further improved by combining time interval and access time information.

For vector hidden dimension, we studied the influence of vector hidden dimension on IPE integral, and increased the dimension from 5 to 50. Figure2 shows the mapping values of different dimensions of the dataset. The evaluation model is stable within the range of [30,50]. We find that our model is better than other algorithms even if the dimension is less than 15. The results also confirm the superiority of our evaluation model.

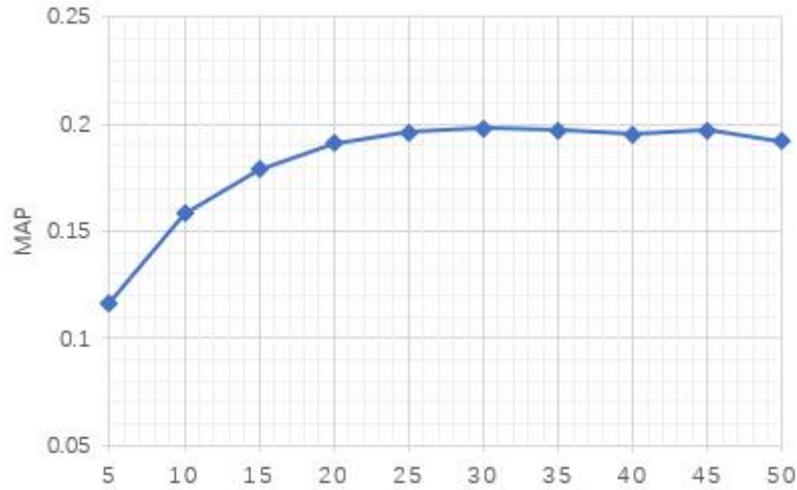

Figure2 Evaluating the impact of model dimensions on performance

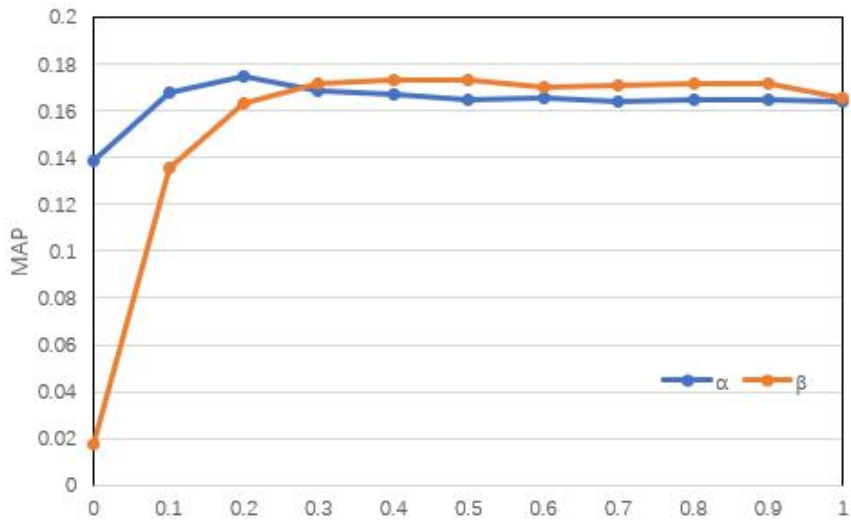

Figure3   Performance at different α and β for fixed α= 0.3 and β= 0.2, respectively

For additional metadata, we further investigated how adding additional metadata information to the score model affects the accuracy of the evaluation. Table 3.4 shows whether friendship and text description are combined in the dataset. We see that the valuation model performs better by combining additional metadata information. Note that $\alpha$ and $\beta$ in formula (3.17), formula (3.15), and formula (3.14) each determine the meaning of POI and user metadata. The two parameters match here. First we fix $\beta$ =1 to select the optimal value of $\alpha$ . According to this strategy, we put $\alpha$ =0.3 and $\beta$ =0.2 in front of the dataset. Figure 3.8 shows the performance of the evaluation model by changing the values of $\alpha$ and B after setting $\alpha$ =0.3 respectively. $\beta$ =0.2. It is clear that as the value of $\alpha$ or $\beta$ gradually increases to 1, the performance of the valuation model begins to weaken. So the best range of $\beta$ is [0.1,0.3], and the best range of $\alpha$ is [0.3,0.6]. We find the user-related metadata information in the dataset more useful. In general, the experimental results show that the proposed evaluation model can use additional metadata to achieve better evaluation accuracy.

In general, the scoring model can learn user intent well from the user's contextual data to better evaluate the following POI recommendations returned by the edge cloud server.

## 6 Conclusion

In this paper, the privacy protection framework when users initiate location-based service requests in the edge environment is carried out with the priority of user service quality, and its privacy protection approach and the way of evaluating the results returned by the edge server and cloud server are investigated. A privacy protection scheme based on pseudonymity and k-anonymity is designed, and RSA encryption algorithm is used to encrypt the user's private data. Two evaluation models are also proposed: NPE and POE, which combine privacy protection and related methods of destination prediction and evaluation, and can order the services returned by the server to the user to be more in line with the user's needs and can satisfy the user while safeguarding the user's privacy. Based on this, subsequent research directions can be to optimize the decision and response speed of task reproduction while protecting user privacy, and more reasonable resource allocation schemes need to be considered.